\documentclass[fleqn,12pt,twoside]{article}
\usepackage{espcrc1}

\title{ 
Lattice Predictions for Hybrids and
Glueballs.
}
\author{
Craig McNeile
\address{Dept. of Math Sci., University of Liverpool, L69 3BX, UK.} 
}
\begin{document}

\maketitle

\begin{abstract}
I review the results from lattice gauge theory for the 
properties of the light $1^{-+}$ exotic state and $0^{++}$ glueball.
\end{abstract}

\section{Introduction}

QCD is a simple elegant theory, but it is hard to solve it using
anything other than perturbation theory.
Quantities, such as masses, depend on the coupling ($g$) like $M \sim
e^{-1/g^2}$~\cite{Montvay:1994cy}, hence perturbation theory can't be
used to compute the masses of hadrons such as the proton. 
Our inability to solve QCD non-perturbatively makes it hard
to determine basic parameters of the quark sector of the standard model.
For example, the allowed range on the strange quark mass in the particle data
table~\cite{Groom:2000in} is 75 to 170 MeV; a range of almost 100\%.

%
% why important:
% any understanding of confinement
% solving realitic strongly coupled picture
% m \sim exp( - 1/g^2), beyond perturbation theory
%
%%% One of the central aims of the proposed QCD machines is to study
%%% confinement. It would be inconceivable to claim that confinement is
%%% ``understood'', until the nature of the light scalars and exotic meson
%%% spectrum is computed from QCD and the predictions verified by
%%% experiment. To do this will require theoretical tools, such as lattice
%%% QCD calculations, that can compute quantities nonperturbatively.

A particularly good test of our understanding of the
non-perturbative aspects of QCD is to study particles where the gauge
field is excited somehow, and hence playing a more important dynamic
role than in ``standard'' hadrons.  Examples of such particles are
glueballs (particles made out of the gauge fields) and hybrid mesons
($\overline{q}q$ and excited glue). A very good overview of
the interesting issues in hadronic physics, that contrasts the hadron
spectroscopy approach to the study of confinement with the results
from DIS type studies, is the white paper by Capstick et
al.~\cite{Capstick:2000dk}.  The only technique that offers any
prospect of computing masses and matrix elements non-perturbatively,
from first principles, is lattice QCD.  

I review the results
from the lattice for the lightest $0^{++}$ glueball and $1^{-+}$
exotic, as these states are the closest to being
experimentally confirmed. Other recent 
reviews~\cite{Michael:2001qz,Morningstar:2001nu,Bali:2001nc} of
lattice results for hybrids and glueballs focus on different
aspects of the subject.

\section{Lattice QCD calculations}

%%
%%  path integral
%%  (Euclidean space )
%%
Many bound state properties of QCD can be 
determined from the path integral
%%
%% check whether the dagger is at t or 0
%%
\begin{equation}
c(t) \sim \int dU  \int d\psi  \int d\overline{\psi}
\;
\sum_{\underline{x}} O(\underline{0},0) 
O(\underline{x},t)^{\dagger}
e^{-S_F - S_G }
\label{eq:reallyQCD}
\end{equation}
where $S_F$ is the fermion action (some lattice
version of the continuum Dirac action) and $S_G$
is the pure gauge action. 
The path integral in eq.~\ref{eq:reallyQCD} is put on
the computer using a clever finite 
difference formalism~\cite{Montvay:1994cy}, due 
to Wilson, that maintains gauge invariance.
The path integral in eq.~\ref{eq:reallyQCD} is evaluated using
algorithms that are  generalisations of the Monte Carlo methods
used to compute low dimensional integrals.
The physical picture for eq.~\ref{eq:reallyQCD} is that a
hadron is created at time 0, from where it propagates to the time t,
where it is destroyed.
%%
%%
%%
%%  simulation data
%%
The physics from the calculation is extracted using a 
fit model~\cite{Montvay:1994cy}:
\begin{equation}
c(t) = a_0 exp( -m_0 t ) + a_1 exp( -m_1 t ) + \cdots
\label{eq:fitmodel}
\end{equation}
where $m_0$ ($m_1$) is the ground (first excited) state
mass and the dots represent higher excitations.
Although in principle excited state masses can be extracted from a
multiple exponential fit, in practice this is a numerically non-trivial
task, because of the noise in the data from the calculation.
Any gauge invariant combination of quark fields and gauge links can be
used as interpolating operators ($O(\underline{x},t)$) in 
eq.~\ref{eq:reallyQCD}.

The fermion integration can be done exactly in eq.~\ref{eq:reallyQCD}
to produce the fermion determinant.  The determinant describes the
dynamics of the sea quarks. In quenched QCD calculations,
the determinant is set to a constant.
Quenched calculations are roughly 1000 times
cheaper computationally than the calculations that include the
dynamics of the sea quarks. Quenched QCD gives quite a reasonable
description of experiment. For example, the most accurate quenched
calculation of the hadron spectrum, to date, has been completed by the
CP-PACS collaboration~\cite{Kanaya:1998sd}.  From the masses of 11
light hadrons, they conclude that the quenched approximation disagrees
with experiment by at most 11\%. 

In an individual lattice calculation  there are errors from the finite
size of the lattice spacing and the finite lattice volume. State of the art
lattice calculations in the quenched theory, run at a number of
different lattice spacings and physical volumes and extrapolate the
results to the continuum and infinite volume~\cite{Kanaya:1998sd} limit. 
The increased
computational costs of unquenched calculations means that most
calculations are currently done at fixed lattice spacings. One of the most
interesting unquenched calculations is being performed by 
the MILC collaboration~\cite{Bernard:2001av}. 
MILC's calculations include  
2+1 flavours of sea quarks with a lattice spacing of 0.13 fm,
box size of 2.6 fm, and the lightest ratio of the pseudoscalar
to vector mass is 0.4. 
%%%A lattice spacing of 0.05 fm may be
%%%%necessary to take the continuum limit.

\section{Results for glueballs in quenched QCD}

Interpolating operators for glueballs are constructed 
for eq.~\ref{eq:reallyQCD}
from closed loops of gauge links
with the required $J^{PC}$ quantum numbers 
Some highlights of the results are that the lightest glueball is the
$0^{++}$ state with a mass of 1.64(4) GeV~\cite{Teper:1998kw}. 
The next lightest glueball is
$2^{++}$. The ratio of the tensor to scalar glueball mass is
$M_{2^{++}}/M_{0^{++}}$ = 1.42(6)~\cite{Teper:1998kw}.  
The spectrum of glueball states for
other $J^{PC}$ quantum numbers with masses under 4 GeV has been 
comprehensively 
mapped out by Morningstar and Peardon~\cite{Morningstar:1999rf}.

In the real world glueballs will decay to two
mesons, hence they will have a decay width.
Lattice QCD calculations are performed in Euclidean 
space,
for convergence of the path integral in eq.~\ref{eq:reallyQCD}.
The Euclidean nature of lattice calculations makes
the computation of inherently complex quantities such as
decay widths more involved~\cite{Michael:1989mf}.

The GF11 lattice group computed the decay widths for the
decay of the $0^{++}$ glueball to two
pseudoscalars~\cite{Sexton:1995kd} to be 108(28) MeV. 
Although the error bar is only statistical, it is
encouraging that the width was small relative to the 
mass, so the $0^{++}$ glueball may exist as a well defined
state. The decay widths for individual meson pairs~\cite{Burakovsky:1998zg} 
did not
agree with the predictions from the ``flavour democratic'' assumption.

The experimental situation~\cite{Groom:2000in} 
for light $0^{++}$ scalars is
very interesting, because there are too many states to put into
SU(3) nonets, as other particles with 
different $J^{PC}$ quantum numbers,
such as the pseudoscalars, can be.
The hadrons: $f_0(1370)$ ,$f_0(1500)$ , and $f_0(1710)$,
have masses close to the mass of the quenched scalar  $0^{++}$ glueball.

In full QCD interpolating operators with $0^{++}$ can be constructed
from quarks and ante-quarks, such as $\overline{\psi}\psi$.  In full QCD, the
pure glue $0^{++}$ operators will mix with the fermionic $0^{++}$
operators. If the mixing is very strong, then the final $0^{++}$
masses will have little to do with the quenched glueball masses.

Weingarten and Lee~\cite{Lee:1999kv} studied the effect of mixing
between the glueball and $\overline{\psi}\psi$ states in quenched QCD.
They measured the correlation between the $0^{++}$ glueball states
and $\overline{\psi}\psi$ states in eq.~\ref{eq:reallyQCD}. The results
were expressed as a mixing matrix
\begin{equation}
\left(
\begin{array}{cc}
m_g           & E(s)         \\
E(s)          & m_{\sigma}(s)  \\
\end{array}
\right)
\label{eq:mixing}
\end{equation}
%%%
%%%
where $m_g$ is the glueball mass, $m_{\sigma}(s)$ is the mass of the 
non-singlet $0^{++}$ state at strange, and $ E(s)$ is the mixing energy.
Weingarten and Lee measured: $m_g$  = 1648(58) MeV, 
$m_{\sigma}(s)$  = 1322(42) MeV,
and $E(s)$ = 61(58) MeV. The qualitative picture that emerges is that
the $f_0(1710)$ is ``mostly'' $0^{++}$ glueball, and the 
$f_0(1500)$ is ``mostly'' $ \overline{s} s$. 
It is not clear whether $f_0(1500)$ being $\overline{s} s$
is consistent with its decay width~\cite{Close:2001ga}.

The  mixing energy $E(s)$ has large lattice spacing errors.
For example at a lattice spacing of $a^{-1} \sim$ 1.2 GeV,
the Weingarten and Lee~\cite{Lee:1999kv} 
result is $E(s) \sim$ 0.36 GeV. This has been
checked by another group's 
result~\cite{McNeile:2000xx} of $E(s) \sim$ 0.44 GeV.

The analysis of Weingarten and Lee~\cite{Lee:1999kv} 
depends on the
$0^{++}$ states being well defined in quenched QCD.
Eichten~\cite{Bardeen:2001jm} et al. 
have shown that there is a problem with the 
non-singlet $0^{++}$ correlator in quenched QCD.
The problem can be understood using quenched chiral 
perturbation theory. The non-singlet $0^{++}$ propagator
contains an intermediate state of $\eta'-\pi$. The removal of fermion
loops in quenched QCD has a big effect on the $\eta'$ propagator.
The result is that a ghost state contributes to the scalar correlator,
that makes the expression in eq.~\ref{eq:fitmodel} inappropriate to 
extract masses from the calculation.
Eichten et al.~\cite{Bardeen:2001jm} predict that the ghost state will
make the $a_0$ mass increase as the quark mass is reduced below 
a certain point.
This behaviour was
observed by Weingarten and Lee~\cite{Lee:1999kv} 
for small box sizes (L $\le$ 1.6 fm) for
quark masses below strange. It is not clear how the problem
with the non-singlet $0^{++}$ correlator in the quenched approximation 
effects the results of Weingarten and Lee~\cite{Lee:1999kv}, however
their most important results come from  masses above
the strange quark mass where the ghost
diagram  will make a smaller contribution that may be negligible

%%Recently, there has been a number of attempts to compute
%%the glueball spectrum using the 
%%ADS supergravity duality (see Brower at al.~\cite{Brower:2000rp}, that
%%also contains references to earlier work). The glueball spectrum in the 
%%large N (number of colours) limit of QCD is obtained.
%%Teper and Lucini~\cite{Lucini:2001ej} have systematically studied 
%%the glueball spectrum in the large N limit of QCD, by calculating the 
%%glueball spectrum for $N$ = 2,3,4 and 5. The dependence of the 
%%glueball spectrum on N is weak.

%%There are lighter $0^{++}$ states, such as the $f_0(980)$ and the
%%$a_0(980)$, and the enigmatic $f_0(400-1200)$. The $f_0(980)$ and
%%$a_0(980)$ states are considered by some people to be kaon molecules
%%or 4 quark states, although there are other opinions.  There has been
%%some recent work by Alford and Jaffe on 4 quark states.

\section{Results for glueball masses in two flavour QCD}

The Weingarten and Lee~\cite{Lee:1999kv} 
analysis predicted that the mixing of
the $0^{++}$ glueball and $\overline{\psi} \psi $ states is small. 
Parts of their calculation have been criticised
in~\cite{McNeile:2000xx},
however, the
problems with the non-singlet $0^{++}$ correlator~\cite{Bardeen:2001jm}
 in the quenched QCD
will make further progress in mixing in the quenched QCD difficult.

A lattice QCD calculation that included the dynamics of the sea quarks
should just reproduce the physical spectrum of $0^{++}$ states. Some
insight into the composition of individual $0^{++}$ states, such as
whether a physical particle couples to $\overline{\psi}\psi$ or pure glue
operators, could be studied by looking at the effect of decreasing the
sea quark mass. For very heavy sea quark masses the theory is more
like quenched QCD, where glueballs are distinct from $\overline{\psi}\psi$
operators.

%%It is not trivial to extract the masses of singlet
%%mesons from quenched calculations, as the correlators
%%depend on the fermion loops.

Hart and Teper~\cite{Hart:2001fp} 
found that the ratio of the $0^{++}$ glueball mass
in $n_f = 2$ QCD to the quenched QCD result
was: $M_{n_f=2}^{0^{++}}/ M_{quenched}^{0^{++}}$
= $0.84 \pm 0.03$ at a fixed lattice spacing of 0.1 fm. 
The $n_f$ = 2 results~\cite{Hart:2001fp}  for the mass of the $2^{++}$ 
were consistent with the quenched value.
As the 
lattice spacing
dependence of the mass of the singlet $0^{++}$ state in two flavour QCD and
quenched QCD could be different, a definitive result will only come
after a continuum extrapolation of the unquenched masses.
In  quenched QCD~\cite{Teper:1998kw}, 
the difference between
the continuum extrapolated mass of $0^{++}$ glueball mass and 
the mass at 0.1 fm is of the order of 200 MeV. This is the same
magnitude of the mass splittings between the masses of the 
experimentally observed particles $f_0(1500)$ and $f_0(1710)$.
Although the current results for singlet $0^{++}$ states are 
starting to be interesting, the lattice spacing 
used in unquenched calculations
must be reduced before 
direct contact can be made to phenomenology.

The mass of the $0^{++}$ glueball on the UKQCD data 
set are just degenerate with the mass of 
two pions~\cite{Hart:2001fp}. As the mass of the 
sea quarks is reduced, two pion states may effect the physics
of singlet $0^{++}$ states.

\section{Results for light $1^{-+}$ exotic mesons}

The quark model predicts the 
charge conjugation (C = $(-1)^{L+S}$ ) 
and parity (P=  $(-1)^{L+1}$ ) of a meson
with spin $S$ and orbital angular momentum $L$.
States with quantum numbers
not predicted by the quark model, such as:
$J_{exotic}^{PC}$ = $1^{-+}$, $0^{+-}$, $2^{+-}$, $0^{--}$
are known as exotics~\cite{Burnett:1990aw}.  
Exotic states are allowed by QCD.
Morningstar and Peardon~\cite{Morningstar:1999rf} claim
that there are no glueballs with exotic quantum numbers
with masses less than 4 GeV.

There are a number of different possibilities for the structure of an
exotic state. An exotic signal could be: a hybrid meson, that is
a quark and anti-quark with excited glue, or bound state of two quarks and
two anti-quarks ($\overline{\psi}\overline{\psi}\psi\psi$).

One possible interpolating operator~\cite{Bernard:1997ib},
that can be used in eq.~\ref{eq:reallyQCD},
for a  hybrid $1^{-+}$ particle is
\begin{equation}
O_{1^{-+}} (\underline{x} , t ) = \overline{\psi}(\underline{x},t)
\gamma_j F_{ij} (\underline{x},t)
\psi (\underline{x},t)
\label{eq:interponemp}
\end{equation}
where $F$ is the QCD field strength tensor.
If $F$ is removed from eq.~\ref{eq:interponemp}, the operator
creates the $\rho$  particle.
In this formalism a gauge invariant interpolating operator, for any
possible exotic hybrid particle or four particle state can be
constructed. 
The dynamics then determines whether the resulting state
has a narrow decay width, hence it can be detected experimentally. In the
large $N_c$ (number of colours)
limit~\cite{Burnett:1990aw,Cohen:1998jb} both exotic hybrid mesons and
non-exotic mesons have widths that are small compared to their
masses.
%%%

There have not been many new calculations of the mass of the light
$1^{-+}$ hybrid  recently.  All the results from the various lattice QCD
calculations, by UKQCD~\cite{Lacock:1997ny,Lacock:1996vy},
MILC~\cite{Bernard:1997ib,McNeile:1998cp} and
SESAM~\cite{Lacock:1998be} are essentially consistent  with the mass
of the $1^{-+}$ state around $1.9(2)$ GeV~\cite{Michael:2001qz}.  The
interpolating operators used to create the exotic meson states in the
MILC calculations~\cite{Bernard:1997ib} are different to those used in
the UKQCD~\cite{Lacock:1997ny} and SESAM
simulations~\cite{Lacock:1998be}, hence giving confidence that the
systematic errors are under control.  The results for the hybrid
masses reported by Lacock and Schilling~\cite{Lacock:1998be}, include
some effects from dynamical sea quarks.
The recent results for the $1^{-+}$ mass from calculations that used an
asymmetric~\cite{Mei:2002ip} lattice in time are consistent with the older results.

There are a number of experimental candidates for light $1^{-+}$
states~\cite{Groom:2000in}.
The E852 collaboration have
reported~\cite{Adams:1998ff} a signal for $1^{-+}$ state
around $1.6$ GeV.  
There is also an experimental signal for a $1^{-+}$ 
state at $1.4$ GeV~\cite{Groom:2000in}. 
%%
%%%See
%%%Page for the arguments that this state really is a
%%%hybrid meson and not a $\overline{Q}\overline{Q}QQ$ state.

There has been some recent work~\cite{Thomas:2001gu} on the quark mass
dependence of the $1^{-+}$ states. The lattice calculations are
usually done at large quark masses and the results extrapolated to the
physical quark masses.  The conclusion of~\cite{Thomas:2001gu} was
that the inclusion of the decay of the hybrid in the quark mass
dependence of the exotic mass could reduce the final answer by 100
MeV. The predictions
 in~\cite{Thomas:2001gu} will be tested as the quark masses used in
lattice calculations are reduced. The $1^{-+}$ state at 1.4 GeV seems
low
relative to the lattice results.

%%
%%Clearly the agreement between the possible experimental signals for
%%the $1^{-+}$ states and the lattice results is very poor.
%%% at two and
%%%three lattice sigma.  
%%The errors on the lattice results for $1^{-+}$
%%states are large relative to the errors on $\overline{Q}Q$ states. To
%%quantify the disagreement between experiment and lattice results the
%%systematic errors on the lattice simulation results should be reduced.
%%In particular the masses of the quarks used in the lattice simulation
%%should be reduced.  

It is possible that the states seen experimentally
are really $\overline{\psi}\overline{\psi}\psi\psi$ states, in which case the
operators used in the lattice simulations (eq.~\ref{eq:interponemp}) 
might not couple strongly to
them. Alford and Jaffe~\cite{Alford:2000mm} 
studied $\overline{\psi}\overline{\psi}\psi\psi$ operators
with $J^{PC}$ = $0^{++}$ in a recent lattice calculation.
The motivation was to gain insight into states such as the 
$f_0(980)$ that some people believe is not a $\overline{\psi}\psi$ 
meson, but a $\overline{\psi}\overline{\psi}\psi\psi$ state.
A similar lattice calculation could in principle be done for the 
$J^{PC}$ = $1^{-+}$ exotic.

%%
%% insight diquarks Bielefeld
%% Test models 
%%
%% Title: Diquarks...
%%   First-Author: Anselmino
%%   Journal-ref: Rev.Mod.Phys. 65 (1993) 1199 
%%
%%The results from lattice QCD also provide insight into the underlying
%%dynamics of light hadrons. Lattice QCD simulations can test the
%%various assumptions made in models of the QCD dynamics.  For example
%%there are a number of models of exotic states based on the idea of a
%%bound diquark anti-diquark
%%pair~\cite{Uehara:1996hc}.  A
%%critical assumption in diquark models is that two quarks actually do
%%cluster to form a diquark. This assumption has been tested in a
%%lattice gauge theory simulation by the Bielefeld
%%group~\cite{Hess:1998sd}, where they found no deeply bound diquark
%%state in Landau gauge.

%%
%%  UKQCD strange - light is 120 MeV
%%

%%  decay rates
%%

To definitely identify a particle requires both the calculation of the
mass as well as the decay widths.  There has been very little work on
strong  decays on the lattice.  
The most obvious hadronic process to study using lattice
gauge theory is the $\rho \rightarrow \pi\pi$ decay, however there
have only been a few attempts to calculate the $g_{\rho \pi \pi}$
coupling~\cite{Gottlieb:1984rh,Loft:1989sy}.
Michael discusses the problems with the formalism for
hadronic decays on the lattice~\cite{Michael:1989mf}. 

In the static quark limit the exotic states on the lattice are
described by adiabatic potentials. The ground state of the static
potential ($A_{1g}$) is the familiar Coulomb plus linear potential. The
excited potential ($E_u$) is a very flat potential, that can be used
with Schr\"{o}dinger's equation to predict the spectrum of heavy-heavy
hybrids~\cite{Michael:2001qz}. 
 UKQCD~\cite{McNeile:2002az} have investigated the
de-excitation of the $E_u$ potential to the $A_{1g}$ potential by the
emission of a light quark loop. In the real world, the decays
would correspond to $1^{-+} \rightarrow \chi_b \eta$ and $1^{-+}
\rightarrow \chi_b S$ with $S$ a scalar and $\eta$ a pseudoscalar. 
The decay width of $1^{-+}
\rightarrow \chi_b \eta$ and $1^{-+} \rightarrow \chi_b S$ transitions
were less than 1 MeV and around 80 MeV respectively.  The various
approximations in the static limit mean that these widths have no
direct relevance to experiment.

The MILC collaboration~\cite{Bernard:1997ib} have
investigated the 
mixing between the operator in eq.~\ref{eq:interponemp}  and
the operator ($\pi \otimes a_1$) eq.~\ref{eq:QQQQ}.
%%%
\begin{equation}
\overline{\psi}^a \gamma_5 \psi^{a} 
\overline{\psi}^b \gamma_5 \gamma_i \psi^{b} 
\label{eq:QQQQ}
\label{eq:fourquark}
\end{equation}
that has the quantum numbers $1^{-+}$.  This type of correlator is
part of the calculation required to compute 
the decay width of the $1^{-+}$ state to
$\rho$, and $a_1$.  The more complicated part  is to
use eq.~\ref{eq:fourquark} in eq.~\ref{eq:reallyQCD} requires some
clever numerical work.

\section{Conclusions}

The glueball spectrum from quenched QCD is essentially
complete. The key issues now are to quantify the mixing of the
glueball states with $\overline{\psi}\psi$ operators and to determine decay
widths.

%%\section{Acknowledgements}
%%%
%%I thank Chris Michael and Doug Toussaint for 
%%discussions about exotic mesons.
%%

%%
%%  make sure the references really look like the 
%%  example style sheet
%%

%%%\bibliographystyle{elsart-num}
%%\bibliographystyle{h-elsevier2}
%%\bibliography{conferences,biblio,hybrid}

\begin{thebibliography}{10}

\bibitem{Montvay:1994cy}
I. Montvay and G. Munster,
\newblock Quantum fields on a lattice,
\newblock Cambridge, UK: Univ. Pr. (1994) 491 p. (Cambridge monographs on
  mathematical physics).

\bibitem{Groom:2000in}
Particle Data Group, D.E. Groom et~al.,
\newblock Eur. Phys. J. C15 (2000) 1,
\newblock %%CITATION = EPHJA,C15,1;%%.

\bibitem{Capstick:2000dk}
S. Capstick et~al.,
\newblock Key issues in hadronic physics, 2000, hep-ph/0012238.

\bibitem{Michael:2001qz}
C. Michael,
\newblock Glueballs, hybrid and exotic mesons, 2001, hep-ph/0101287.

\bibitem{Morningstar:2001nu}
C. Morningstar,
%%\newblock Gluonic excitations in lattice qcd: A brief survey, 2001,
\newblock  2001,
  nucl-th/0110074.

\bibitem{Bali:2001nc}
G.S. Bali,
%%%\newblock 'glueballs': Results and perspectives from the lattice,
%%%2001,
\newblock  2001,
  hep-ph/0110254.

\bibitem{Kanaya:1998sd}
K. Kanaya et~al.,
\newblock Nucl. Phys. Proc. Suppl. 73 (1999) 189, hep-lat/9809146,
\newblock %%CITATION = HEP-LAT 9809146;%%.

\bibitem{Bernard:2001av}
C.W. Bernard et~al.,
\newblock Phys. Rev. D64 (2001) 054506, hep-lat/0104002,
\newblock %%CITATION = HEP-LAT 0104002;%%.

\bibitem{Teper:1998kw}
M.J. Teper,
\newblock Glueball masses and other physical properties of su(n) gauge theories
  in d = 3+1: A review of lattice results for theorists, 1998, hep-th/9812187.

\bibitem{Morningstar:1999rf}
C.J. Morningstar and M.J. Peardon,
\newblock Phys. Rev. D60 (1999) 034509, hep-lat/9901004,
\newblock %%CITATION = HEP-LAT 9901004;%%.

\bibitem{Michael:1989mf}
C. Michael,
\newblock Nucl. Phys. B327 (1989) 515.

\bibitem{Sexton:1995kd}
J. Sexton, A. Vaccarino and D. Weingarten,
\newblock Phys. Rev. Lett. 75 (1995) 4563, hep-lat/9510022.

\bibitem{Burakovsky:1998zg}
L. Burakovsky and P.R. Page,
\newblock Phys. Rev. D59 (1999) 014022, hep-ph/9807400,
\newblock %%CITATION = HEP-PH 9807400;%%.

\bibitem{Lee:1999kv}
W.J. Lee and D. Weingarten,
\newblock Phys. Rev. D61 (2000) 014015, hep-lat/9910008,
\newblock %%CITATION = HEP-LAT 9910008;%%.

\bibitem{Close:2001ga}
F.E. Close and A. Kirk,
\newblock Eur. Phys. J. C21 (2001) 531, hep-ph/0103173,
\newblock %%CITATION = HEP-PH 0103173;%%.

\bibitem{McNeile:2000xx}
C. McNeile and C. Michael,
\newblock Phys. Rev. D63 (2001) 114503, hep-lat/0010019,
\newblock %%CITATION = HEP-LAT 0010019;%%.

\bibitem{Bardeen:2001jm}
W. Bardeen et~al.,
\newblock Phys. Rev. D65 (2002) 014509, hep-lat/0106008,
\newblock %%CITATION = HEP-LAT 0106008;%%.

\bibitem{Hart:2001fp}
A. Hart and M. Teper,
\newblock Phys. Rev. D65 (2002) 034502, hep-lat/0108022,
\newblock %%CITATION = HEP-LAT 0108022;%%.

\bibitem{Burnett:1990aw}
T.H. Burnett and S.R. Sharpe,
\newblock Ann. Rev. Nucl. Part. Sci. 40 (1990) 327.

\bibitem{Bernard:1997ib}
MILC, C. Bernard et~al.,
\newblock Phys. Rev. D56 (1997) 7039, hep-lat/9707008.

\bibitem{Cohen:1998jb}
T.D. Cohen,
\newblock Phys. Lett. B427 (1998) 348, hep-ph/9801316.

\bibitem{Lacock:1997ny}
P. Lacock et~al.,
\newblock Phys. Lett. B401 (1997) 308, hep-lat/9611011.

\bibitem{Lacock:1996vy}
P. Lacock et~al.,
\newblock Phys. Rev. D54 (1996) 6997, hep-lat/9605025.

\bibitem{McNeile:1998cp}
C. McNeile et~al.,
\newblock Nucl. Phys. Proc. Suppl. 73 (1999) 264, hep-lat/9809087,
\newblock %%CITATION = HEP-LAT 9809087;%%.

\bibitem{Lacock:1998be}
 P. Lacock and K. Schilling,
\newblock Nucl. Phys. Proc. Suppl. 73 (1999) 261, hep-lat/9809022,
\newblock %%CITATION = HEP-LAT 9809022;%%.

\bibitem{Mei:2002ip}
Z.H. Mei and X.Q. Luo,
\newblock 2002, hep-lat/0206012.

\bibitem{Adams:1998ff}
 G.S. Adams et~al.,
\newblock Phys. Rev. Lett. 81 (1998) 5760,
\newblock %%CITATION = PRLTA,81,5760;%%.

\bibitem{Thomas:2001gu}
A.W. Thomas and A.P. Szczepaniak,
\newblock Phys. Lett. B526 (2002) 72, hep-ph/0106080,
\newblock %%CITATION = HEP-PH 0106080;%%.

\bibitem{Alford:2000mm}
M.G. Alford and R.L. Jaffe,
\newblock Nucl. Phys. B578 (2000) 367, hep-lat/0001023,
\newblock %%CITATION = HEP-LAT 0001023;%%.

\bibitem{Gottlieb:1984rh}
S. Gottlieb et~al.,
\newblock Phys. Lett. 134B (1984) 346,
\newblock %%CITATION = PHLTA,134B,346;%%.

\bibitem{Loft:1989sy}
R.D. Loft and T.A. DeGrand,
\newblock Phys. Rev. D39 (1989) 2692,
\newblock %%CITATION = PHRVA,D39,2692;%%.

\bibitem{McNeile:2002az}
C. McNeile, C. Michael and P. Pennanen,
\newblock Phys. Rev. D65 (2002) 094505, hep-lat/0201006,
\newblock %%CITATION = HEP-LAT 0201006;%%.

\end{thebibliography}

\end{document}